\documentclass[conference]{IEEEtran}
\IEEEoverridecommandlockouts
\usepackage{cite}
\usepackage{amsmath,amssymb,amsfonts} 
\usepackage{graphicx}
\usepackage{textcomp} 
\def\BibTeX{{\rm B\kern-.05em{\sc i\kern-.025em b}\kern-.08em
    T\kern-.1667em\lower.7ex\hbox{E}\kern-.125emX}} 
\usepackage{url}
\usepackage{hyperref}  
\usepackage{times}
\usepackage{soul}
\usepackage[utf8]{inputenc}
\usepackage{caption}
\usepackage{amsmath}
\usepackage{amsthm}
\usepackage{booktabs} 
\usepackage[switch]{lineno}
\usepackage{makecell}
\usepackage{adjustbox}
\usepackage{mathtools}
\usepackage{subcaption}
\usepackage{lipsum}
\usepackage{wrapfig}
\usepackage{graphicx}
\usepackage{tablefootnote} 
\PassOptionsToPackage{table}{xcolor}
\usepackage{xcolor} 
\usepackage[linesnumbered,ruled,vlined]{algorithm2e}

\title{VoicePrompter: Robust Zero-Shot Voice Conversion with Voice Prompt and Conditional Flow Matching}
 
\author{ 
\IEEEauthorblockN{Ha-Yeong Choi and Jaehan Park}% 
\IEEEauthorblockA{\textit{Gen AI Lab, KT R$\&$D}}
} 
\begin{document}
\maketitle
 
\begin{abstract} 
Despite remarkable advancements in recent voice conversion (VC) systems, enhancing speaker similarity in zero-shot scenarios remains challenging. This challenge arises from the difficulty of generalizing and adapting speaker characteristics in speech within zero-shot environments, which is further complicated by mismatch between the training and inference processes. To address these challenges, we propose VoicePrompter, a robust zero-shot VC model that leverages in-context learning with voice prompts. VoicePrompter is composed of (1) a factorization method that disentangles speech components and (2) a DiT-based conditional flow matching (CFM) decoder that conditions on these factorized features and voice prompts. Additionally, (3) latent mixup is used to enhance in-context learning by combining various speaker features. This approach improves speaker similarity and naturalness in zero-shot VC by applying mixup to latent representations. Experimental results demonstrate that VoicePrompter outperforms existing zero-shot VC systems in terms of speaker similarity, speech intelligibility, and audio quality. Our demo is available at \url{https://hayeong0.github.io/VoicePrompter-demo/}.  
\end{abstract} 

\vspace{-0.4cm}
\section{Introduction}
\label{sec:intro}
Zero-shot voice conversion (VC) systems \cite{casanova2022yourtts,lee2023hiervst,wang2023lm,lee2023hierspeech++,neekhara2023selfvc,li2024sef,kameoka2024voicegrad,anastassiou2024seed} have gained significant attention with the advancement of deep generative models. In particular, diffusion-based VC models \cite{popov2022diffusionbased} have demonstrated high performance in zero-shot speaker adaptation through iterative sampling processes. Recent works, such as Diff-HierVC \cite{choi23d_interspeech} and DDDM-VC \cite{choi2024dddm}, have further improved zero-shot VC performance by adopting source-filter disentanglement and disentangled denoising processes. NaturalSpeech 3 \cite{ju2024naturalspeech} enhanced voice style transfer performance by disentangling the speech into timbre, prosody, content, and residual details. However, diffusion-based models are limited by slow inference speed due to their iterative sampling. Additionally, these models are vulnerable to noisy speech, often generating noisy sound when conditioned on global style embedding extracted from noisy target speech.

Meanwhile, recent advancements in text-to-speech models have shifted from global style conditioning \cite{jia2018transfer,skerry2018towards,wang2018style,lee2021multi} to voice prompting methods \cite{wang2023neural,le2023voicebox,liu2024generative,wang2024speechx,eskimez2024e2} for target speaker adaptation. VALL-E \cite{wang2023neural} was the first to adopt in-context learning for speaker adaptation by concatenating the audio codec into the input sequences. Similarly, VoiceBox \cite{le2023voicebox} was trained using a masking and infilling speech, where speech was generated by infilling masked input sequences with conditional flow matching (CFM). Speech generation with prompting mechanisms can endow the model with in-context learning capability, enabling them to follow the style of a given voice prompt. However, VC models with voice prompts have not yet been thoroughly investigated, primarily due to the challenges of speech disentanglement.  

In this paper, we present VoicePrompter, a robust zero-shot VC model with in-context learning ability using voice prompt. We first adopt a diffusion transformer with CFM as the backbone model. For speech perturbation, we train the model to estimate the vector field based on features extracted by a speech disentangle encoder, augmented with latent mixup. Then, the model is trained by masking sequences and infilling speech to emerge in-context learning ability. The results demonstrate that it is essential to improve the robustness by prompting the target voice when infilling speech from the augmented speech presentation using mixup. By guiding the target voice style with prompts during conversion, our model achieves better speaker similarity compared to recent powerful baselines. 
%For information perturbation, we train the model to estimate the vector field from the source-filter representation, augmented by latent mixup. 
Our main contributions are summarized as follows:
\begin{itemize}
\item We propose VoicePrompter, a robust zero-shot VC system that leverages in-context learning with voice prompts to achieve high speaker similarity. 
\item We improve the robustness of VC by incorporating latent mixup and speech infilling approaches.
\item Thanks to the integration of the CFM and adaLN-sep within the DiT backbone, VoicePrompter achieves successful VC and high audio quality in a single step.
\item  The results show that our model outperforms recent powerful baselines in terms of speaker similarity, speech intelligibility, and audio quality.
\end{itemize}

\begin{figure*}[!t]
\centering
\centerline{\includegraphics[width=0.98\textwidth]{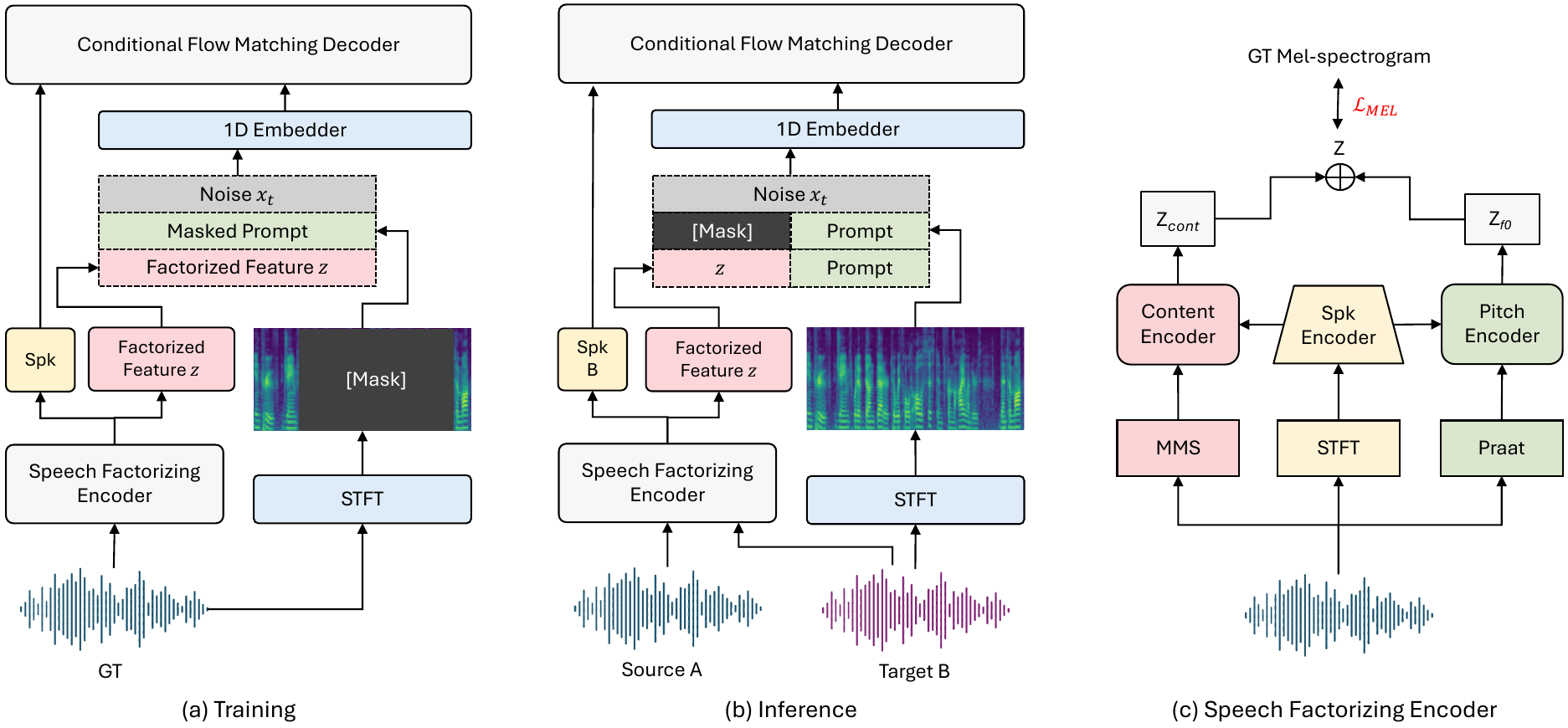}}\vspace{-0.1cm}
\caption{Overall architecture of VoicePrompter. (a) Training phase; (b) Inference phase; (c) Speech Factorizing Encoder}
\label{fig:main}\vspace{-0.3cm}
\end{figure*} 
 
\section{Voice Prompter} 
\label{sec:method}  
In this section, we introduce our proposed system, VoicePrompter. As depicted in Fig. \ref{fig:main}, our model consists of two main components: (1) a speech factorizing encoder that effectively disentangles and embeds the input speech, and (2) a DiT-based CFM decoder that conditions on factorized speech features and voice prompts. The details are described in the following subsection.     
 
\subsection{Speech Factorizing Encoder}  
\subsubsection{Content Encoder} 
To extract linguistic information from input audio, we utilize the seventh layer of a pre-trained MMS \cite{pratap2023scaling} model. Given that MMS embeddings include acoustic information, we apply signal perturbation to the input audio to isolate linguistic information independent of speaker characteristics. The extracted MMS embeddings are then modeled alongside speaker information using an 8-layer WaveNet-based \cite{oord2016wavenet} content encoder.

\subsubsection{Pitch Encoder}
For pitch extraction, we employ Praat \cite{jadoul2018introducing} to obtain F0 values. These extracted F0 values are embedded according to the encoder's hidden layer configuration, and the embedded pitch information is processed through a temporal bottleneck layer. Subsequently, a pitch encoder, utilizing the same WaveNet architecture as the content encoder, models the pitch information.

\subsubsection{Speaker Encoder}
We extract speaker information by applying spectral feature extraction on the Mel-spectrogram using 1D convolutional layers. Temporal features are refined with a Conv1dGLU layer, and long-term dependencies are captured through multi-head attention. A final 1D convolutional layer generates the style representation, which is then used for speaker adaptation across all encoders and decoders.
 
\renewcommand{\textfloatsep}{5pt}  
\begin{algorithm}
\caption{Compute CFM Loss}
\label{algo_loss}
\SetInd{0em}{1em}
\KwIn{Factorized speech feature $z$, voice prompt $p$, speaker emb $e_{spk}$, Vector field estimator $v_{\theta}$}
\KwOut{Loss value $\mathcal{L}_{\text{CFM}}$}
\SetKwFunction{ComputeCFMLoss}{ComputeCFMLoss}
\SetKwProg{Fn}{Function}{:}{} 
\Fn{\ComputeCFMLoss{$x_1$, $z$, $p$, $e_{spk}$}}{  
    Sample $t \sim \mathcal{U}(0, 1)$\;
    Sample $x_0 \sim \mathcal{N}(0, \mathbf{I})$\; 
    \textit{where $x_0$ has the same shape as $x_1$}
    
    Compute $\phi_{t}(x_1) \gets (1 - (1 - \sigma_{\text{min}}) t) x_0 + t x_1$\;
    Compute $u_t^{\text{OT}} \gets x_1 - (1 - \sigma_{\text{min}}) x_0$\;
    Estimate $u_t^{\text{pred}}
, m_{idx} \gets v_{\theta}( x_1, \phi_{t}(x_1), z, e_{spk}, p, t) $\;
    Compute $\mathcal{L}_{\text{CFM}} \gets \text{MSE}(u_t^{\text{pred}}
 \odot m_{idx}, u_t^{\text{OT}}
 \odot m_{idx})$\;
    \Return $\mathcal{L}_{\text{CFM}}$\; 
} 
\While{training}{
    Take batch and sample $x_1$ from training data\; 
    $\mathcal{L}_{\text{CFM}} \gets \ComputeCFMLoss(x_1, z, p, e_{spk})$\;
    Update model weights:
    $\theta \gets \theta - \eta \nabla_{\theta} \mathcal{L}_{\text{CFM}}$\;
} 
\end{algorithm}    

\subsection{Conditional Flow Matching Decoder}
To generate high-quality Mel-spectrograms, we employ a conditional flow matching \cite{lipman2022flow,tong2023conditional} structure utilizing an optimal transport (OT) path. FM starts with a noise sample $x_0$ drawn from a standard Gaussian distribution and learns the time-conditioned transformation $\phi_t$ that maps it to the target sample $x_1$, with this flow controlled by an ordinary differential equations (ODEs). The time-conditioned vector field $u_t$ can be chosen as an OT path, and the corresponding vector field is estimated by the vector field estimator network $v_\theta$. In this process, the network is conditioned on the factorized speech feature $z$, voice prompt $p$, and speaker embedding $e_{\text{spk}}$ to predict the vector field. The computation of the CFM loss is detailed in Algorithm \ref{algo_loss}.  

\subsection{Voice Prompt for In-Context Learning} 
Drawing inspiration from previous work \cite{le2023voicebox,eskimez2024e2} that leveraged masking strategies for in-context learning, we explore a method to incorporate direct target voice prompts into the VC task. Unlike prior approaches, which used only encoded feature from the encoder as conditioning, we introduce a method where the input speech is masked by 70-100$\%$ and utilized alongside the decoder's embedding as conditioning input. Specifically, we adopt the same masking strategy as VoiceBox \cite{le2023voicebox}, computing the CFM loss only on the masked segments. 
In the inference phase, as shown in Fig. \ref{fig:main}-(b), we perform an in-filling task where the masked portions are predicted using the source content information and the factorized feature $z$,  which encodes the target's timbre.  
 
\begin{figure}[t]
    \centering {\includegraphics[width=0.95\columnwidth]{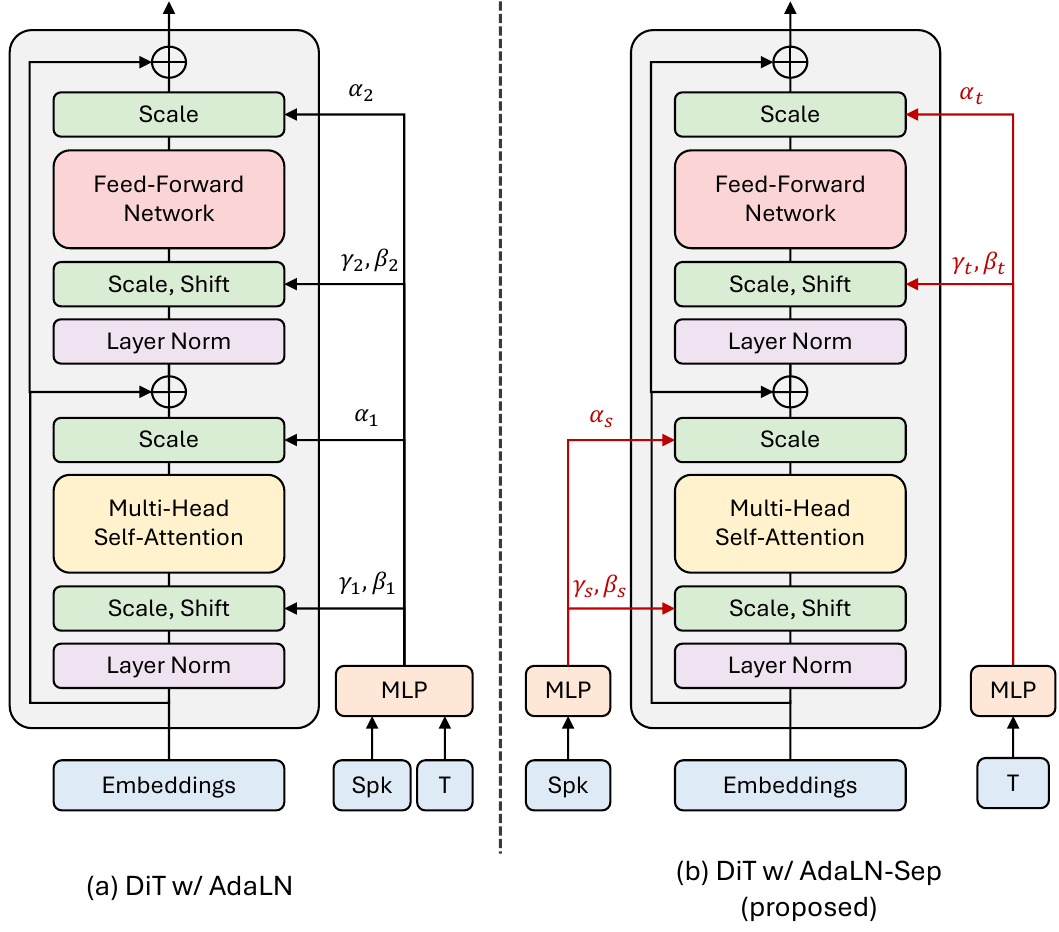}}
    \caption{Comparison of (a) the original DiT block with adaLN-Zero and (b) the proposed DiT block with adaLN-Sep.} 
    \label{fig:dit} 
\end{figure}
 
\subsection{DiT with AdaLN-Sep}  
We employ DiT as the backbone for the CFM decoder and introduce adaLN-Sep, a conditioning method derived from a modification of adaLN-Zero. Fig. \ref{fig:dit}-(a) depicts the adaLN-Zero approach, which demonstrated superior performance in previous DiT \cite{peebles2023scalable} research by exploring various methods of integrating conditioning into transformer blocks. The adaLN-Zero can be described as follows: 
\begin{equation} 
\scalebox{0.8}{$
    \textit{AdaLN-Zero} (h, c) =  {\alpha}_{c} \odot \left(  {\gamma}_{c} \cdot LN(h) +  {\beta}_{c} \right), 
$}
\end{equation}   
The adaLN-Zero block accelerates training by zero-initializing the scaling parameters at the end of each residual block and applying dimension-wise scaling parameters before the residual connections. However, the conventional adaLN-zero has a limitation in that it processes the conditioning information together, which does not fully reflect the independent characteristics of each feature. To address this and enhance both speaker similarity and training efficiency, we propose adaLN-Sep, which separates speaker and time embeddings, as illustrated in Fig. \ref{fig:dit}-(b). This method independently integrates each conditioning information into the transformer block, and adaLN-Sep is defined as follows: 
\begin{equation}\vspace{-0.3cm}
\scalebox{0.85}{$
\textit{AdaLN-Sep} (h, s, t) = 
\begin{cases} 
    {\alpha}_{s} \odot \left( {\gamma}_{s} \cdot \textit{LN}(h) + {\beta}_{s} \right), & \\ \text{         (for SA Block)} \\
    {\alpha}_{t} \odot \left( {\gamma}_{t} \cdot \textit{LN}(h) + {\beta}_{t} \right), & \\ \text{         (for FFN Block)}
\end{cases}
$}\vspace{-0.2cm}
\end{equation}
\subsection{Latent Mixup} 
Although the speaker adaptation performance has improved through the integration of the voice prompt and the powerful DiT backbone network, train-inference mismatch problem still remains in zero-shot VC. Following \cite{choi2024dddm}, during the training phase, we perform latent mixup by randomly combining representations from different speakers to perturb the speech components. Specifically, latent mixup is conducted on 50$\%$ of the batch size. When mixup is not applied, the process can be represented as follows:
\begin{equation}
 {E}_{cont}(z_{\text{cont}, x}, z_{\text{spk}, x}) + {E}_{F0}(z_{\text{F0}, x}, z_{\text{spk}, x}) = \hat{z}  \,
\end{equation}
where $E_{cont}$ denotes the content encoder, $E_{F0}$ denotes the F0 encoder, and $z_{spk, x}$ represents the style information of speaker $x$. In the case where mixup is applied, the formulation is expressed as follows:
\begin{equation}
{E}_{cont}(z_{\text{cont}, x}, z_{\text{spk}, y}) + {E}_{F0}(z_{\text{F0}, x}, z_{\text{spk}, y}) = \hat{z}_{mix} \,
\end{equation}
where $z_{cont, x}$ and $z_{F0, x}$ refer to the content and F0 information of speaker $x$, and $z_{spk, y}$ represents the style information of speaker $y$. We use $\hat{z}_{mix}$ as a condition for the CFM decoder, and $\hat{z}$ is employed for the encoder's reconstruction. This strategy allows us to disentangle and embed the relevant factors more robustly, ultimately leading to improved generalization in the zero-shot VC task.
  
\begin{table*}[t] 
\caption{Zero-shot VC results from VCTK dataset. We used the official checkpoints provided by the authors for the baseline.} 
\label{table:zero-shotVC_result} 
\centering
 \resizebox{1\textwidth}{!}{
\begin{tabular}{l|l|cc|cccc|ccc}
\toprule
Method &Model & Dataset & Hours & CER ($\downarrow$)  & WER ($\downarrow$) & EER ($\downarrow$) & SECS ($\uparrow$) & UTMOS ($\uparrow$) & NMOS ($\uparrow$) & SMOS ($\uparrow$) \\
\midrule
&  GT & - & - &   0.21 & 2.17 & - & - & 4.04 & 4.03$\pm$0.06 & 4.25$\pm$0.04  \\
\midrule 
Codec &  FACodec (NS 3) \cite{ju2024naturalspeech}  & Librilight & 60k & \textbf{0.54}  &  \textbf{2.58}  & 6.03 & 0.869 & 3.28 &3.62$\pm$0.05 & 3.92$\pm$0.05\\  
\midrule   
& DiffVC \cite{popov2022diffusionbased}& LT-460  & 0.2k & 6.86 & 13.77 & 9.25 & 0.826 &  3.49 & 3.43$\pm$0.05  & 3.24$\pm$0.06 \\
Diffusion    &  Diff-HierVC \cite{choi23d_interspeech}   & LT-460 & 0.2k & 0.83 & 3.11 & 3.29 & 0.861 &  3.34& 3.83$\pm$0.04 & 4.01$\pm$0.05 \\
& DDDM-VC \cite{choi2024dddm} & LT-460 &0.2k & 1.77 & 4.35 & 6.49 & 0.858  &  3.40& 3.88$\pm$0.05  & 3.93$\pm$0.05 \\   
\midrule 
% 295k / 590
\rowcolor{gray!10} CFM  & VoicePrompter (Ours) & LT-460 &0.2k & 0.76 & 2.97 & 2.28 & 0.865  & \textbf{3.85} & 3.93$\pm$0.05  & \textbf{4.13$\pm$0.05} \\
\rowcolor{gray!10} &  VoicePrompter (Ours)  & LT-960 &0.5k & 0.62  & \textbf{2.58} & \textbf{1.84} & \textbf{0.872}  & 3.77  &   \textbf{3.97$\pm$0.04}  & 4.10$\pm$0.04 \\
\bottomrule
\end{tabular}}  \vspace{-0.3cm}
\end{table*} 
 
\begin{table}[t]
\centering
\caption{Results of ablation study (LT-460)} 
\label{ablationT}
\resizebox{1.0\columnwidth}{!}{
\begin{tabular}{l|cccccc}
\toprule
 Method &  CER ($\downarrow$)  & WER ($\downarrow$) &  EER ($\downarrow$) & SECS ($\uparrow$) &  UTMOS ($\uparrow$)  \\
\midrule
\rowcolor{gray!10} \textbf{VoicePrompter}  & 0.76 & 2.97  & 2.28 & 0.864& 3.85 \\ 
\midrule 
w/o Mixup  & 0.77 & 2.98 & 3.00 & 0.858 & 3.78 \\
w/o Prompt & 0.77 & 2.81& 5.75 & 0.853  & 3.81 \\
w/o Prompt, Mixup &  0.58 & 2.50 & 8.57 & 0.830  & 3.78 \\ 
\midrule
w/o AdaLN-Sep  & 0.80 & 2.88  & 3.26 & 0.854 & 3.75 \\ 
\bottomrule
\end{tabular}
} 
\end{table} 
 
\section{Experiment and Result}  
\subsection{Experimental Setup}
\textbf{Dataset } We trained our model using the multi-speaker LibriTTS \cite{zen2019libritts}, specifically the $\textit{train-clean-100}$ and $\textit{train-clean-360}$ subsets, which include 245 hours of speech from 1,151 speakers. For validation, we used the $\textit{dev-clean}$ subset. To evaluate zero-shot VC, we selected random sentences from the VCTK \cite{veaux2017superseded}. \\ 
\textbf{Preprocessing   } We resampled the audio to 16,000 Hz using the Kaiser-best algorithm from torchaudio \cite{yang2021torchaudio}. The Mel-spectrogram was generated with a hop size of 320, a window size of 1280, an FFT size of 1280, and a bin size of 80. \\
\textbf{Implementation Details   } We trained the model for 300K steps with a batch size of 64 on two NVIDIA A100 GPUs, and applied the same setup for training the ablation models. The learning rate was set to $2\times10^{-4}$, and the AdamW optimizer was used. For the vocoder, we trained BigVGAN \cite{lee2023bigvgan} on LibriTTS, adapting it to our 16 kHz Mel settings. %, utilizing four A6000 GPUs for training. 
 
\subsection{Zero-shot Voice Conversion} 
 We conduct various subjective and objective evaluation on the zero-shot VC task with four strong VC baseline: DiffVC \footnote{\url{https://github.com/huawei-noah/Speech-Backbones/tree/main/DiffVC}}, Diff-HierVC \footnote{\url{https://github.com/hayeong0/Diff-HierVC}}, DDDM-VC \footnote{\url{https://github.com/hayeong0/DDDM-VC}}, and NaturalSpeech (NS) 3's VC model, FACodec \footnote{\url{https://huggingface.co/amphion/naturalspeech3_facodec}}. 
 Each model was evaluated using official checkpoints, with a consistent sampling of 6 steps applied to all baselines except NS 3 for fair comparison. For subjective evaluation, we conduct the naturalness (NMOS) and similarity mean opinion score (SMOS), and UTMOS \cite{saeki22c_interspeech}. For objective evaluation, we utilized four key metrics: character error rate (CER), word error rate (WER), equal error rate (EER), and speaker encoder cosine similarity (SECS). To evaluate the accuracy of intelligibility, we used Whisper-large-v2 \cite{radford2023robust} to measure CER and WER, and evaluated the EER using an automatic speaker verification model. We also calculated SECS with Resemblyzer. The results in Table \ref{table:zero-shotVC_result} show that our model delivers strong performance in terms of speaker similarity, speech clarity, and overall audio quality. 
 
\begin{table}[h!] 
  \caption{Conversion results based on sampling steps} 
  \centering 
  \resizebox{0.9\columnwidth}{!}{
    \begin{tabular}{c|c|ccccc}
    \toprule
    Model & Timestep & CER & WER & EER & SECS & UTMOS \\
    \midrule  
      & 1 & 0.52 & 2.70 & 3.25 & 0.861 & 3.85 \\
      & 2 & 0.54 & 2.70 & 2.27 & 0.862 & 3.92  \\
VoicePrompter & 3 & 0.61 & 2.83 & 2.25 & 0.864 & 3.91 \\
      & 4 & 0.62 & 2.83 & 2.28 & 0.864 & 3.89 \\
      & 6 &  0.76 & 2.97 & 2.28 & 0.865  & 3.85  \\
      & 10 & 0.86 & 3.10 & 2.50 & 0.865 & 3.80 \\ 
    \bottomrule
\end{tabular}} 
\label{table:ablation_ts}
\end{table}  

\subsection{Ablation Study}
We conducted ablation studies for latent mixup, voice prompt, and AdaLN-Sep to demonstrate the effectiveness of the proposed methods. We first followed the mixup of DDDM-VC \cite{choi2024dddm} to perturb the speaker information before fed to the CFM decoder. However, as shown in Table \ref{ablationT}, while latent mixup improved speaker similarity, it led to a decrease in audio quality. This reduction in quality is likely due to the perturbed representation, which can affect the model's robustness. To address this issue, we utilized voice prompts along with latent mixup to guide the voice information and enhance the model's robustness during training. Table \ref{ablationT} shows that using both mixup and voice prompts mechanism significantly improves the performance in terms of speaker similarity and audio quality. Furthermore, AdaLN-Sep could enhance the adaptation performance by conditioning speaker and time embeddings separately, and we also observed that AdaLN-Sep accelerates the training process.  
 
\subsection{Sampling Steps}
We compared the performance of VoicePrompter according to sampling steps. We found that our model could convert the speech even with a single step generation. However, increasing the sampling steps consistently increase the speaker similarity in terms of SECS. Although the UTMOS of each result showed a similar score, we found that increasing the sampling step could improve the perceptual quality. We have added the audio samples based on the sampling steps on the demo page\footnote{\url{https://hayeong0.github.io/VoicePrompter-demo/}}. 

\begin{table}[t] 
  \caption{Details of model variants} 
  \centering 
  \resizebox{1\columnwidth}{!}{
    \begin{tabular}{l|c|cccc}
    \toprule
    Model & Params. & Layers & Hidden & MLP & Heads \\
    \midrule   
     VoicePrompter & 155M & 12 & 768  & 3072 & 12   \\
     VoicePrompter-S &  38M & 8  & 384  & 1536 & 8    \\ 
    \bottomrule
\end{tabular}}  
  \label{table:ablation_size}
\end{table}  

\begin{table}[t]
\centering
\caption{Results based on different model size (LT-960)}  
\label{ablation_size_results}
\resizebox{1\columnwidth}{!}{
\begin{tabular}{l|cccccc}
\toprule
 Method  & CER ($\downarrow$) & WER ($\downarrow$)& EER ($\downarrow$)& SECS ($\uparrow$) & UTMOS ($\uparrow$)   \\
\midrule   
VoicePrompter & 0.62 & 2.58 & 1.84 & 0.872 & 3.77 \\  
VoicePrompter-S & 0.63 & 2.58 & 2.59 & 0.870 & 3.60 \\  
\bottomrule
\end{tabular}
}  
\end{table} 

\subsection{Scaling Down Model Size}
We scale down model size to evaluate the robustness of our structure. The details of model hyperparameter are described in TABLE \ref{table:ablation_size}. Table \ref{ablation_size_results} reveals that VoicePrompter-S still showed lower CER and WER compared to baseline models. Furthermore, both models has better speaker similarity than other baselines. The results also demonstrated that increasing the model size could enhance the capability for voice style transfer. Moreover, scaling up model size could significantly improve the audio quality. In future work, we will further scale up both model size and data size for better generalization.  
 
\section{Conclusion}
In this paper, we proposed \textit{VoicePrompter}, a zero-shot VC model designed to enhance in-context learning capabilities through the voice prompts. Our model adopts a DiT as its backbone, incorporating adaLN-sep, and estimates vector fields using a flow matching conditioned on factorized speech features. This design allowed the model to achieve robust speaker adaptation performance. Notably, we introduced a voice prompt method that combined latent mixup with sequence masking, which significantly improved the robustness. Experimental results demonstrated that using target voice prompts during inference process could maximize speaker similarity. 
\textit{VoicePrompter} showed outstanding performance in zero-shot domain, proving that prompting techniques offered new possibilities in VC tasks. The findings suggested that prompting could greatly enhance perceptual performance in VC and highlighted the potential of high-quality backbone models to maintain superior audio quality.  
  
\section{Acknowledgements}
We'd like to thank Sang-Hoon Lee for valuable discussions.
\bibliographystyle{IEEEbib}
\bibliography{vp.bib} 
\end{document}